\documentclass[amsmath,amssymb,%showkeys,
showpacs,twocolumn, nobibnotes,aps,prd]{revtex4-1}

\usepackage{graphicx,amsmath,amssymb}
\usepackage{mathrsfs,eufrak}
\usepackage{dcolumn,color}
\usepackage{bm}
\begin{document}
%%%%%%%%%%%%%%%%%%%%%%%%%%%%%%%%%%%%%%%%%%%%%%%%%%%%%%%%%%%%%%%%%%%%%%%%%%%%%%%

\title{Two-impurity-entanglement generation by electron scattering in zigzag phosphorene nanoribbons}

%%%%%%%%%%%%%%%%%%%%%%%%%%%%%%%%%%%%%%%%%%%%%%%%%%%%%%%%%%%%%%%%%%%%%%%%%%%%%%%
\author{M. Amini}
\author{M. Soltani}\email{mo.soltani@sci.ui.ac.ir}
\author{E. Ghanbari-Adivi}
\author{M. Sharbafiun}
%%%%%%%%%%%%%%%%%%%%%%%%%%%%%%%%%%%%%%%%%%%%%%%%%%%%%%%%%%%%%%%%%%%%%%%%%%%%%%%
\affiliation{Department of Physics, Faculty of Sciences, University
of Isfahan, Isfahan 81746-73441, Iran}
%%%%%%%%%%%%%%%%%%%%%%%%%%%%%%%%%%%%%%%%%%%%%%%%%%%%%%%%%%%%%%%%%%%%%%%%%%%%%%%
\begin{abstract}
In this paper, we investigate how two on-side doped impurities with
net magnetic moments in an edge chain of a zigzag phosphorene
nanoribbon~(zPNR) can be entangled by scattering of the traveling
edge-state electrons. To this end, in the first step, we employ the
Lippmann-Scwinger equation as well as the Green's function approach
to study the scattering of the free traveling electrons from two
magnetic impurities in a one-dimensional tight-binding chain. Then,
following the same formalism, that is shown that the behavior of two
on-side spin impurities in the edge chain of a zPNR in responding to
the scattering of the edge-state traveling electrons is very similar
to what happens for the one-dimensional chain. In both cases,
considering a known incoming wave state, the reflected and
transmitted parts of the final wave state are evaluated
analytically. Using the obtained results, the related partial
density matrices and the reflection and transmission probabilities
are computable. Negativity as a measure of the produced entanglement
in the final state is calculated and the results are discussed. Our
theoretical model actually proposes a method, which is perhaps
experimentally performable to create the entanglement in the state
of the impurities .
\end{abstract}
%%%%%%%%%%%%%%%%%%%%%%%%%%%%%%%%%%%%%%%%%%%%%%%%%%%%%%%%%%%%%%%%%%%%%%%%%%%%%%%
%%%%%%%%%%%%%%%%%%%%%%%%%%%%%%%%%%%%%%%%%%%%%%%%%%%%%%%%%%%%%%%%%%%%%%%%%%%%%%%
%%%%%%%%%%%%%%%%%%%%%%%%%%%%%%%%%%%%%%%%%%%%%%%%%%%%%%%%%%%%%%%%%%%%%%%%%%%%%%%
\maketitle
%%%%%%%%%%%%%%%%%%%%%%%%%%%%%%%%%%%%%%%%%%%%%%%%%%%%%%%%%%%%%%%%%%%%%%%%%%%%%%%
%%%%%%%%%%%%%%%%%%%%%%%%%%%%%%%%%%%%%%%%%%%%%%%%%%%%%%%%%%%%%%%%%%%%%%%%%%%%%%%
%%%%%%%%%%%%%%%%%%%%%%%%%%%%%%%%%%%%%%%%%%%%%%%%%%%%%%%%%%%%%%%%%%%%%%%%%%%%%%%
%%%%%%%%%%%%%%%%%%%%%%%%%%%%%%%%%%%%%%%%%%%%%%%%%%%%%%%%%%%%%%%%%%%%%%%%%%%%%%%
%%%%%%%%%%%%%%%%%%%%%%%%%%%%%%%%%%%%%%%%%%%%%%%%%%%%%%%%%%%%%%%%%%%%%%%%%%%%%%%
%%%%%%%%%%%%%%%%%%%%%%%%%%%%%%%%%%%%%%%%%%%%%%%%%%%%%%%%%%%%%%%%%%%%%%%%%%%%%%%
\section{Introduction~\label{Sec01}}
Since the realization of phosphorene~\cite{Liu2014}, as an atomic
layer of phosphorus, it has attracted great attention due to its
physical properties and possible
applications~\cite{Carvalho2016,Ling2015}. This new specimen of~2D
material is a layered crystal of phosphorus atoms which are
covalently bonded with three nearest neighbors via $sp^3$
hybridization to form a puckered~2D honeycomb structure. It is this
unique property of anisotropy together with a large direct band gap
that makes phosphorene a desirable candidate material for different
applications with specific electronic, mechanical, thermal, and
transport
features~\cite{Neto2014,Qiao2014,Peeters2014,Guinea2014,Yang2014,Katsnelson2015,Asgari2015}.
Moreover, phosphorene like the other 2D nanomaterials can be
patterned into phosphorene nanoribbons~(PNR) which can be fabricated
with lithography and plasma etching of
black-phosphorus~\cite{Feng2015,Maity2016,Fang2017}. The phosphorene
ribbon with zigzag edges, which is called zigzag phosphorene
nanoribbon~(zPNR), shows degenerate quasi-flat bands in the middle
of the gap that separates the valence and conduction
bands~\cite{Ezawa2014}. Furthermore, due to the existence of such a
gap that protects the quasi-flat band, the quantum transport of
these localized edge sates are found to be, in some sense, like a
quasi-one dimensional chain both numerically and
analytically~\cite{Ezawa2014,Asgari2018}.\par
%%%%%%%%%%%%%%%%%%%%%%%%%%%%%%%%%%%%%%%%%%%%%%%%%%%%%%%%%%%%%%%%%%%%%%%%%%%%%%%%
On the other hand, one of the interesting aspects of modern quantum
mechanics is to share quantum information and create entanglement
between the components of a quantum-scale system. One of the ways to
this end is to use the scattering phenomena. In a number of pervious
studies in this field, adopting a free electron model, the
entanglement generation created due to one-dimensional scattering of
electrons from Kondo and Heisenberg impurities has been
theoretically
investigated~\cite{Ghanbari2013,Ghanbari2015,Ghanbari2016}. Also, in
one of these studies, it was shown that the Klein tunneling effect
of the traveling electrons in a graphene sheet can leads to a
correlation between the transmitted and the reflected electrons and
also between the quantum-dot spin qubits fixed in the graphene
nanoribbons~\cite{Ghanbari2016}. However, the differences between
the electronic spectrum of the graphene and phosphorene
nanostructures can motivate development of such studies to
nanostructures based on phosphorus. In fact, one of the significant
aspects of the edge states in zPNR might be the fingerprints of
these states on the entanglement generation between two localized
magnetic impurities on the edges through the scattering of electrons
by such impurities. In the present work, scattering of the ballistic
electrons by the quantum-dot spin qubits fixed at the edges of a
zPNR is investigated theoretically. To this end, since the
phospherene edge states in a zigzag nanoribbon exhibit like a
one-dimensional tight-binding model, we employed the general
scattering theory based on the Green function approach and the
Lippmann-Schwinger  equation to investigate the entanglement
creation due to the scattering of electrons from a one-dimensional
tightly bonded subsystem. In the second step, we assume that two
spin impurities are fixed at the edge sites of phosphorene and the
outlined model is developed to scattering of electrons from these
impurities leading to entanglement generation. In order to show the
similarity of the behavior of the phosphorene edge states with a
one-dimensional  tightly bonded system, the results of the
calculations performed for both cases are compared, showing the good
consistency of the results.\par
%%%%%%%%%%%%%%%%%%%%%%%%%%%%%%%%%%%%%%%%%%%%%%%%%%%%%%%%%%%%%%%%%%%%%%%%%%%%%%%
The significance of this study is that it introduces the considered
case as a physical one-dimensional system which can be possibly used
to produce entanglement between magnetic impurities
experimentally.\par
%%%%%%%%%%%%%%%%%%%%%%%%%%%%%%%%%%%%%%%%%%%%%%%%%%%%%%%%%%%%%%%%%%%%%%%%%%%%%%%
The paper is organized as follows. In section~\ref{Sec02}, we study
the scattering of the free traveling electrons from two spin
impurities doped into a one-dimensional chain. In this section, the
tight-binding Hamiltoninan and the applied scattering approach are
explained. In section~\ref{Sec03}, we generally introduce the
outlined model and the used formalism to calculate the transmission
coefficient for scattering of electrons from edges of zPNRs.
Section~\ref{Sec04} is devoted to discussion on the obtained
results. Finally, we wrap up the paper with summary and concluding
remarks in section~\ref{Sec05}.
%%%%%%%%%%%%%%%%%%%%%%%%%%%%%%%%%%%%%%%%%%%%%%%%%%%%%%%%%%%%%%%%%%%%%%%%%%%%%%%
%%%%%%%%%%%%%%%%%%%%%%%%%%%%%%%%%%%%%%%%%%%%%%%%%%%%%%%%%%%%%%%%%%%%%%%%%%%%%%%
%%%%%%%%%%%%%%%%%%%%%%%%%%%%%%%%%%%%%%%%%%%%%%%%%%%%%%%%%%%%%%%%%%%%%%%%%%%%%%%
%%%%%%%%%%%%%%%%%%%%%%%%%%%%%%%%%%%%%%%%%%%%%%%%%%%%%%%%%%%%%%%%%%%%%%%%%%%%%%%
%%%%%%%%%%%%%%%%%%%%%%%%%%%%%%%%%%%%%%%%%%%%%%%%%%%%%%%%%%%%%%%%%%%%%%%%%%%%%%%
%%%%%%%%%%%%%%%%%%%%%%%%%%%%%%%%%%%%%%%%%%%%%%%%%%%%%%%%%%%%%%%%%%%%%%%%%%%%%%%%
\section{entanglement generation in a one-dimensional tight-binding Chain \label{Sec02}}
\begin{figure}[t]
\centering
\includegraphics[scale=0.7]{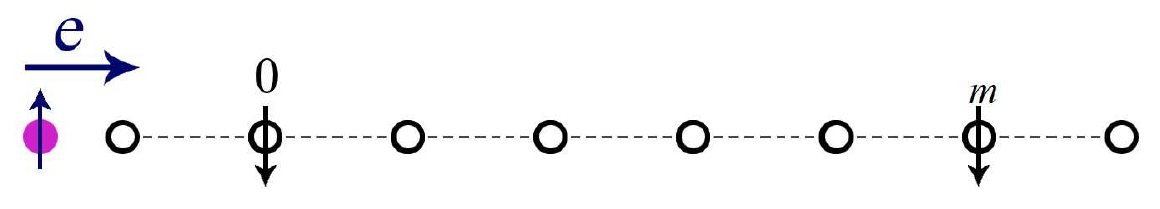}
\caption{Scattering of a free traveling electron by two on-site spin
impurities localized at a distance of $m$ from each other in a
one-dimensional tight-binding chain. } \label{Fig01}
\end{figure}

In this section, we study the entanglement generation between two
on-site spin impurities localized in a one-dimensional
tightly-bonded atomic chain due to the scattering of free electrons.
As is schematically shown in Fig.~\ref{Fig01}, it is assumed that a
free electron moving through the chain impinges on two magnetic
impurities which are fixed at $m$ distance of each other.  So the
electron spin can be seen as a mediator between the spin of the
impurities and the scattering process can lead to entanglement
production in their spin quantum state. Through the calculations,
the Heisenberg operators are used to describe the interactions
between the involved spins.\par
%%%%%%%%%%%%%%%%%%%%%%%%%%%%%%%%%%%%%%%%%%%%%%%%%%%%%%%%%%%%%%%%%%%%%%%%%%%%%%%
%%%%%%%%%%%%%%%%%%%%%%%%%%%%%%%%%%%%%%%%%%%%%%%%%%%%%%%%%%%%%%%%%%%%%%%%%%%%%%%
Without loss of generality, it is assumed that the impurities are
localized at sites $0$ and $m$, respectively. So, the Hamiltonian of
the system reads
%%%%%%%%%%%%%%%%%%%%%%%%%%%%%%%%%%%%%%%%%%%%%%%%%%%%%%%%%%%%%%%%%%%%%%%%%%%%%%%
\begin{equation}
\label{EQ01} H = \sum_{i}^{} t c^{\dagger}_i c_{i+1} + hc + {\hat
V},
\end{equation}
%%%%%%%%%%%%%%%%%%%%%%%%%%%%%%%%%%%%%%%%%%%%%%%%%%%%%%%%%%%%%%%%%%%%%%%%%%%%%%%
where summation runs over all lattice sites and $hc$ stands for
Hermitian conjugate. $t$ is the hopping integral between nearest
neighbors, $c^\dagger_i~(c_i)$ is the creation~(annihilation)
operator of an electron at site $i$ and ${\hat V}$ is the
interaction potential due to the presence of the impurities.\par
%%%%%%%%%%%%%%%%%%%%%%%%%%%%%%%%%%%%%%%%%%%%%%%%%%%%%%%%%%%%%%%%%%%
In the face of the impurities, the electron matter wave is partially
reflected and transmitted. The reflection and transmission
amplitudes can be evaluated using the so-called transition matrix
approach. For a typical scattering potential of ${\hat V}$, the
transition operator ${\hat T}$ is defined as
%%%%%%%%%%%%%%%%%%%%%%%%%%%%%%%%%%%%%%%%%%%%%%%%%%%%%%%%%%%%%%%%%%%%%%%%%%%%%%%
\begin{equation}\label{EQ02}
\hat{T} = \hat{V} \big( 1 + \hat{G}_E \hat{V} + \hat{G}_E
\hat{V}\hat{G}_E \hat{V} + ...\big) = \hat{V} \big( 1 - \hat{G}_E
\hat{V} \big)^{-1},
\end{equation}
%%%%%%%%%%%%%%%%%%%%%%%%%%%%%%%%%%%%%%%%%%%%%%%%%%%%%%%%%%%%%%%%%%%%%%%%%%%%%%%
where $\hat{G}_E$ is the Green's operator of the defect-free system.
Here, both $\hat{T}$ and $\hat{G}_E$ are dependent on the energy of
the system, $E$, but for brevity we refuse to display it
explicitly.\par
%%%%%%%%%%%%%%%%%%%%%%%%%%%%%%%%%%%%%%%%%%%%%%%%%%%%%%%%%%%%%%%%%%%%%%%%%%%%%%%
For a one-dimensional tightly bonded system the matrix elements of
$\hat{G}_E$ in the site basis are given by a closed-form expression
as~\cite{Economou1990}:
%%%%%%%%%%%%%%%%%%%%%%%%%%%%%%%%%%%%%%%%%%%%%%%%%%%%%%%%%%%%%%%%%%%%%%%%%%%%%%%
\begin{equation}
\label{EQ03} G_E(m;m') = \langle m | \hat{G}_E | m' \rangle =
{e^{ik_0|m-m'|}\over 2 i t  \sin k_0},
\end{equation}
%%%%%%%%%%%%%%%%%%%%%%%%%%%%%%%%%%%%%%%%%%%%%%%%%%%%%%%%%%%%%%%%%%%%%%%%%%%%%%%
in which $m$ and $m'$ are the numbers labeling the sites and their
corresponding basis $|m\rangle$ and $|m'\rangle$ are obtained by
acting the creation operators $c^\dagger_m$ and $c^\dagger_{m'}$ on
the vacuum ground state. Also $k_0$ is given by $k_0=\cos^{-1}(E/2
t)$.
\par
%%%%%%%%%%%%%%%%%%%%%%%%%%%%%%%%%%%%%%%%%%%%%%%%%%%%%%%%%%%%%%%%%%%%%%%%%%%%%%%
The scattering potential $\hat V$, due to the presence of the doped
spin impurities in the lattice, is in explicit form of
%%%%%%%%%%%%%%%%%%%%%%%%%%%%%%%%%%%%%%%%%%%%%%%%%%%%%%%%%%%%%%%%%%%%%%%%%%%%%%%
\begin{equation}
\label{EQ04} \hat{V} = U \big[ ({\bf S}_1 \cdot {\bf S}_2)
c_0^\dagger c_0 + ({\bf S}_1 \cdot {\bf S}_3) c^\dagger_m c_m \big],
\end{equation}
%%%%%%%%%%%%%%%%%%%%%%%%%%%%%%%%%%%%%%%%%%%%%%%%%%%%%%%%%%%%%%%%%%%%%%%%%%%%%%%
${\bf S}_1$ is the dimensionless spin operator of the incident
electron, ${\bf S}_2$ and ${\bf S}_3$ are the same for the
impurities, and $U$ is the impurity potential in the system.\par
%%%%%%%%%%%%%%%%%%%%%%%%%%%%%%%%%%%%%%%%%%%%%%%%%%%%%%%%%%%%%%%%%%%%%%%%%%%%%%%
In a representation space including the involved spins as well as
the site states $|0\rangle$ and $|m\rangle$, this potential can be
represented as a squared $16\times 16$ matrix, while the spin
interactions are $4\times 4$ matrices in their corresponding
subspaces. For example, in the computational basis the spin
interaction  ${\bf S}_1\cdot {\bf S}_2$ is given by
%%%%%%%%%%%%%%%%%%%%%%%%%%%%%%%%%%%%%%%%%%%%%%%%%%%%%%%%%%%%%%%%%%%%%%%%%%%%%%%
\begin{equation}
\label{EQ05} {\bf S}_1\cdot {\bf S}_2 = {1\over 4} \vert t \rangle
\langle t \vert - {3\over 4}\vert s \rangle \langle s \vert
\end{equation}
%%%%%%%%%%%%%%%%%%%%%%%%%%%%%%%%%%%%%%%%%%%%%%%%%%%%%%%%%%%%%%%%%%%%%%%%%%%%%%%
where $t$ and $s$ refer to triplet~(symmetric) and
singlet~(asymmetric) spin states, respectively. Clearly, the
explicit matrix form of this interaction is
%%%%%%%%%%%%%%%%%%%%%%%%%%%%%%%%%%%%%%%%%%%%%%%%%%%%%%%%%%%%%%%%%%%%%%%%%%%%%%%
\begin{equation}
\label{EQ06} {\bf S}_1\cdot {\bf S}_2 = \begin{bmatrix}
{1\over 4} &          0     &         0      &      0 \\
    0      & -   {1\over 4} &  \ \   {1\over 2} &      0 \\
    0      & \ \ {1\over 2} &   - {1\over 4} &      0 \\
    0      &         0      &         0      & {1\over 4}
\end{bmatrix}.
\end{equation}
%%%%%%%%%%%%%%%%%%%%%%%%%%%%%%%%%%%%%%%%%%%%%%%%%%%%%%%%%%%%%%%%%%%%%%%%%%%%%%%
A similar matrix form can be derived for the other spin interaction,
${\bf S}_1\cdot {\bf S}_3$. Consequently, the interaction potential
cab be written as
%%%%%%%%%%%%%%%%%%%%%%%%%%%%%%%%%%%%%%%%%%%%%%%%%%%%%%%%%%%%%%%%%%%%%%%%%%%%%%%
\begin{equation}
\label{EQ07} \hat{V} = U \big[ ({\bf S}_1\cdot {\bf
S}_2\big)_{4\times4} \otimes {\bf 1}_3  \otimes |0 \rangle \langle
0| + \big({\bf S}_1\cdot {\bf S}_3\big)_{4\times4} \otimes {\bf 1}_2
\otimes | m \rangle \langle m | \big],
\end{equation}
where $ {\bf 1}_2 $ and  ${\bf 1}_3$ are the identity matrices on
the spin spaces of the impurities and $|0 \rangle \langle 0|$ and $|
m \rangle \langle m |$ are the on-site projection operators . The
interaction operator can be also rewritten in a more compact form of
\begin{equation}
\label{EQ08} \hat{V} = {\hat V}_{00} \otimes |0\rangle \langle 0 | +
 {\hat V}_{mm} \otimes  | m \rangle \langle m|,
\end{equation}
with
\begin{equation}
\label{EQ09} \begin{split} {\hat V}_{00}=\big(V_{00}\big)_{8\times8}
& = U ({\bf S}_1\cdot {\bf
S}_2\big)_{4\times4} \otimes {\bf 1}_3,\\
{\hat V}_{mm}=\big(V_{mm}\big)_{8\times8} & = U \big({\bf S}_1\cdot
{\bf S}_3\big)_{4\times4} \otimes {\bf 1}_2. \end{split}
\end{equation}
%%%%%%%%%%%%%%%%%%%%%%%%%%%%%%%%%%%%%%%%%%%%%%%%%%%%%%%%%%%%%%%%%%%%%%%%%%%%%%%
Using this form of the interaction potential, it is an easy practice
to show that transition matrix $\hat T$, given in Eq.~(\ref{EQ02}),
can be represented as
%%%%%%%%%%%%%%%%%%%%%%%%%%%%%%%%%%%%%%%%%%%%%%%%%%%%%%%%%%%%%%%%%%%%%%%%%%%%%%%
\begin{equation}
\label{EQ10} \hat{T}={\hat V} \big(1-\hat{\chi}\big)^{-1},
\end{equation}
%%%%%%%%%%%%%%%%%%%%%%%%%%%%%%%%%%%%%%%%%%%%%%%%%%%%%%%%%%%%%%%%%%%%%%%%%%%%%%%
where $\hat \chi$ is a 16-dimensional square matrix of the following
form
%%%%%%%%%%%%%%%%%%%%%%%%%%%%%%%%%%%%%%%%%%%%%%%%%%%%%%%%%%%%%%%%%%%%%%%%%%%%%%%
\begin{equation}
\label{EQ11} {\hat \chi} = \begin{bmatrix}
\hat{G}_{00} & \hat{G}_{0m}\\
\hat{G}_{m0} & \hat{G}_{mm}
\end{bmatrix} \, \begin{bmatrix}
\hat{V}_{00} & 0\\
0 & \hat{V}_{mm}
\end{bmatrix}.
\end{equation}
%%%%%%%%%%%%%%%%%%%%%%%%%%%%%%%%%%%%%%%%%%%%%%%%%%%%%%%%%%%%%%%%%%%%%%%%%%%%%%%
In the above equation, matrix block  $\hat{G}_{mm'}$ reads
$\hat{G}_{mm'}=G_E(m;m')~{\bf 1}_{spin}$, where $G_E(m;m')$ is given
in Eq.~(\ref{EQ03}) and  identity operator ${\bf 1}_{spin}$ reads
${\bf 1}_{spin}={\bf 1}_1\otimes {\bf 1}_2 \otimes{\bf 1}_3$.\par
%%%%%%%%%%%%%%%%%%%%%%%%%%%%%%%%%%%%%%%%%%%%%%%%%%%%%%%%%%%%%%%%%%%%%%%%%%%%%%%
As is known, the direct spin~(electron-impurity) interaction  can be
modeled as a short range Heisenberg exchange. Accordingly, we
consider the spatial and spinorial spaces in an incoming wave stat
of $| \Psi_{in} \rangle$ with a wave number of $k_0$ for instance as
$| \Psi_{in} \rangle =  \sum\limits_{m}^{} e^{ik_0m} |m \rangle
|\hspace{-1mm}\uparrow\downarrow\downarrow\rangle$.  In the initial
wave state, it is assumed that the incident electron has initially
spin up while both impurities have spin down in the $z$
direction.\par
%%%%%%%%%%%%%%%%%%%%%%%%%%%%%%%%%%%%%%%%%%%%%%%%%%%%%%%%%%%%%%%%%%%%%%%%%%%%%%%
The outgoing wave state,  $| \Psi_{out} \rangle$, can be obtained as
a solution of the Lippmann-Schwinger equation
\begin{equation}
\label{EQ12}
\begin{split}
|\Psi_{out}\rangle & = \big[1+ \hat{G}_E\hat{T} \big] |\Psi_{in}
\rangle\\            &= \big[1+{\hat G}_E{\hat V}
\big(1-\hat{\chi}\big)^{-1}\big] | \Psi_{in} \rangle\\
& =|\Psi_{in} \rangle + \hat{G}_E|\Psi'  \rangle,
\end{split}
\end{equation}
%%%%%%%%%%%%%%%%%%%%%%%%%%%%%%%%%%%%%%%%%%%%%%%%%%%%%%%%%%%%%%%%%%%%%%%%%%%%%%%
in which the auxiliary wave state $|\Psi'  \rangle$ is defined as
$|\Psi'  \rangle = {\hat V} \big(1-\hat{\chi}\big)^{-1} | \Psi_{in}
\rangle$. Considering the matrix form of the interaction potential,
$\hat V$, and the transition matrix, $\hat T$, it is obvious that
$|\Psi'  \rangle$ has a general form of
%%%%%%%%%%%%%%%%%%%%%%%%%%%%%%%%%%%%%%%%%%%%%%%%%%%%%%%%%%%%%%%%%%%%%%%%%%%%%%%
\begin{equation}
\label{EQ13} |\Psi' \rangle =  |s_0 \rangle | 0 \rangle + |s_m
\rangle | m \rangle,
\end{equation}
%%%%%%%%%%%%%%%%%%%%%%%%%%%%%%%%%%%%%%%%%%%%%%%%%%%%%%%%%%%%%%%%%%%%%%%%%%%%%%%
where $| s_0 \rangle$ and $ | s_m \rangle$ are two total spin states
which are derivable analytically.\par
%%%%%%%%%%%%%%%%%%%%%%%%%%%%%%%%%%%%%%%%%%%%%%%%%%%%%%%%%%%%%%%%%%%%%%%%%%%%%%%
Two parts are included in the outgoing state, the reflected part
$|\Psi_R\rangle$ which is detectable at the left side of the
impurities, and the transmitted part, $|\Psi_T\rangle$, detectable
at those right side. Using Eqs.~(\ref{EQ12}) and~(\ref{EQ13}), the
reflected and transmitted wave states are derivable in the final
form of
%%%%%%%%%%%%%%%%%%%%%%%%%%%%%%%%%%%%%%%%%%%%%%%%%%%%%%%%%%%%%%%%%%%%%%%%%%%%%%%
\begin{equation}
\label{EQ14}
\begin{split}
|\Psi_R \rangle & = e^{- ik_0m'} |S_R\rangle |m'\rangle;\qquad {\rm~for} \ m'<0,\\
|\Psi_T \rangle & = e^{+ ik_0m'} |S_T\rangle |m'\rangle;\qquad {\rm~for} \ m'>m,\\
\end{split}
\end{equation}
%%%%%%%%%%%%%%%%%%%%%%%%%%%%%%%%%%%%%%%%%%%%%%%%%%%%%%%%%%%%%%%%%%%%%%%%%%%%%%%
where the reflected and transmitted spin states, $|S_R\rangle$ and
$|S_T\rangle$ are
\begin{equation}
\label{EQ15}
\begin{split}
|S_R\rangle & = [G_{m'0} | s_0 \rangle + G_{m'm} |s_m \rangle]e^{+ik_0m'},\\
|S_T\rangle & =|\hspace{-1mm}\uparrow\downarrow\downarrow\rangle +
[G_{m'0} |s_0\rangle + G_{m'm} |s_m \rangle] e^{-ik_0m'}.
\end{split}
\end{equation}
%%%%%%%%%%%%%%%%%%%%%%%%%%%%%%%%%%%%%%%%%%%%%%%%%%%%%%%%%%%%%%%%%%%%%%%%%%%%%%%
The reflection, transmission and total partial density matrices,
$\rho_{23R}$, $\rho_{23T}$, and $\rho_{23}$ are defined as
\begin{equation}
\label{EQ16}
\begin{split}
\rho_{23R} &  = Tr_1 \big(|S_R\rangle\langle S_R|\big)~/~Tr\big(|S_R\rangle\langle S_R|\big), \\
\rho_{23T} &  = Tr_1 \big(|S_T\rangle\langle S_T|\big)~/~Tr
\big(|S_T\rangle\langle S_T|\big), \\
\rho_{23}  &  = \rho_{23R} + \rho_{23T},
\end{split}
\end{equation}
where $Tr_1(A)$ stands for the trace of $A$ over the electron spin
degree of freedom. These partial density matrices can be used to
evaluate the amount of the created entanglement between the doped
impurities. Also, the reflection and transmission probabilities can
be evaluated using the above quantum states. These issues will be
further discussed in the Results section.
%%%%%%%%%%%%%%%%%%%%%%%%%%%%%%%%%%%%%%%%%%%%%%%%%%%%%%%%%%%%%%%%%%%%%%%%%%%%%%%
%%%%%%%%%%%%%%%%%%%%%%%%%%%%%%%%%%%%%%%%%%%%%%%%%%%%%%%%%%%%%%%%%%%%%%%%%%%%%%%
%%%%%%%%%%%%%%%%%%%%%%%%%%%%%%%%%%%%%%%%%%%%%%%%%%%%%%%%%%%%%%%%%%%%%%%%%%%%%%%
%%%%%%%%%%%%%%%%%%%%%%%%%%%%%%%%%%%%%%%%%%%%%%%%%%%%%%%%%%%%%%%%%%%%%%%%%%%%%%%
%%%%%%%%%%%%%%%%%%%%%%%%%%%%%%%%%%%%%%%%%%%%%%%%%%%%%%%%%%%%%%%%%%%%%%%%%%%%%%%
%%%%%%%%%%%%%%%%%%%%%%%%%%%%%%%%%%%%%%%%%%%%%%%%%%%%%%%%%%%%%%%%%%%%%%%%%%%%%%%
%%%%%%%%%%%%%%%%%%%%%%%%%%%%%%%%%%%%%%%%%%%%%%%%%%%%%%%%%%%%%%%%%%%%%%%%%%%%%%%
\section{entanglement generation in a zigzag phosphorene nanoribbon\label{Sec03}}
In this section,  the phosphorene edge states in a zigzag nanoribbon
are introduced and their corresponding Green's function is
analytically derived. Scattering of electrons from two on-side spin
impurities doped to the zigzag edges of a phosphorene nanoribbon is
considered. The introduced edge states are used to describe the
incident electrons and the derived Green's function is used to
evaluate the amount of the produced entanglement between the
impurities. As will be seen, the approach is very similar to what
was followed in the previous section.\par
\begin{figure*}[t]
\begin{center}
\includegraphics[scale=0.8]{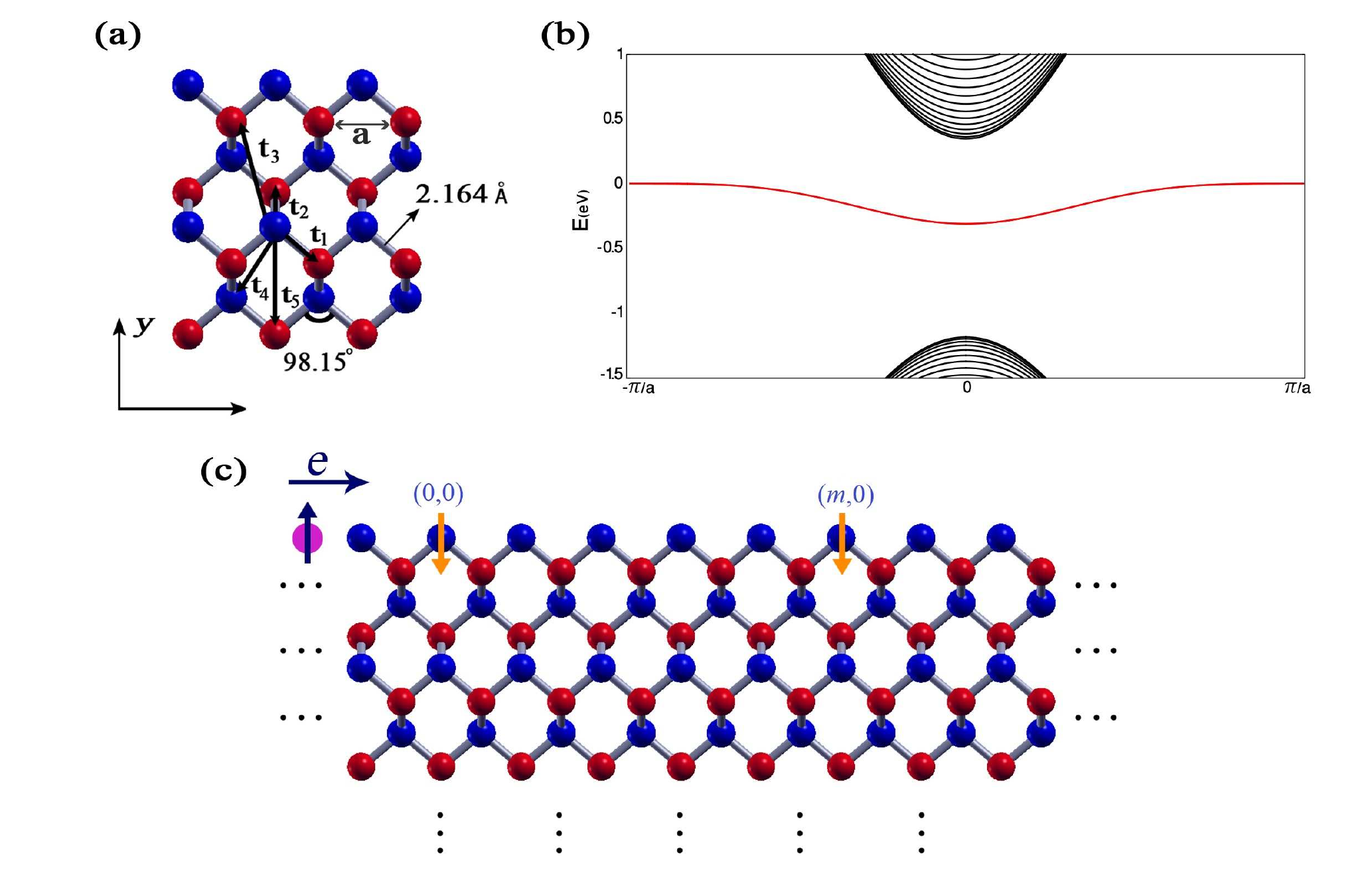}
\caption{(Color online) (a) Schematic of the lattice structure of
phosphorene and hopping integrals $t_{i}$, blue~(A) and red~(B)
colors refer to two types of atoms in the lattice,  (b) band
structure of a phosphorene nanoribbon: there is a considerable gap
between the conduction and valence bands and the degenerate
quasi-flat bands occur separately at the middle of this gap,  (c)
Scattering of the edge-state electrons from the impurities on two
sites of the zigzag edge of a phosphorene nanoribbon, the sites are
labeled as $(0,0)$ and $(m,0)$.}\label{Fig02}
\end{center}
\end{figure*}

%%%%%%%%%%%%%%%%%%%%%%%%%%%%%%%%%%%%%%%%%%%%%%%%%%%%%%%%%%%%%%%%%%%%%%%%%%%%%%%%
%%%%%%%%%%%%%%%%%%%%%%%%%%%%%%%%%%%%%%%%%%%%%%%%%%%%%%%%%%%%%%%%%%%%%%%%%%%%%%%
An impurity-doped infinite zPNR of a given width is schematically
shown in Fig.~\ref{Fig02}. As is seen, the unit cell contains two
atoms labeled $A$ and $B$. The electronic structure of this lattice
is described by a tight-binding Hamiltonian as
%%%%%%%%%%%%%%%%%%%%%%%%%%%%%%%%%%%%%%%%%%%%%%%%%%%%%%%%%%%%%%%%%%%%%%%%%%%%%%%
\begin{equation}
{\hat H} = {\hat H}_0+{\hat V} = \sum_{\langle i,j \rangle} t_{ij}
c_{i}^\dagger c_j + hc + {\hat V}, \label{EQ17}
\end{equation}
%%%%%%%%%%%%%%%%%%%%%%%%%%%%%%%%%%%%%%%%%%%%%%%%%%%%%%%%%%%%%%%%%%%%%%%%%%%%%%%
%%%%%%%%%%%%%%%%%%%%%%%%%%%%%%%%%%%%%%%%%%%%%%%%%%%%%%%%%%%%%%%%%%%%%%%%%%%%%%%
where $\langle i,j \rangle$ stands for the nearest-neighbor index,
$c^\dagger_i~(c_i)$ is the same as introduced previously, $t_{ij}$
is the hopping integral between sites $i$ and $j$, and ${\hat V}$ is
the Hamiltonian due to the presence of the impurity.\par
%%%%%%%%%%%%%%%%%%%%%%%%%%%%%%%%%%%%%%%%%%%%%%%%%%%%%%%%%%%%%%%%%%%%%%%%%%%%%%%
Using the~\textit{ab~initio} method~\cite{Rudenko2014}, it has been
shown that only five hopping parameters are sufficient to describe
the band structure of phosphorene. Referring to Fig.~\ref{Fig02}, we
indicate these parameters for simplicity by $t_1$ to $t_5$. The
corresponding values for these parameters are $t_1 =- 1.220~eV$,
$t_2 = 3.665~eV $, $t_3 = - 0.205~eV$, $t_4 = - 0.105~eV$, and $t_5
= - 0.055~eV$. The interaction term including $t_4$ causes the
particle-hole symmetry breaking in the lattice and it should be kept
in the further simplifications. But, in comparison with $t_1$ and
$t_2$, it will be a good approximation if one neglects the smaller
values of $t_3$ and $t_5$.\par
%%%%%%%%%%%%%%%%%%%%%%%%%%%%%%%%%%%%%%%%%%%%%%%%%%%%%%%%%%%%%%%%%%%%%%%%%%%%%%%
Accordingly, an effective anisotropic honeycomb lattice model was
developed~\cite{Ezawa2014} to analytically describe the electronic
structure of phosphorene. In that model, the terms including $t_1$
and $t_2$ was considered as the tight-binding Hamiltonian with exact
solutions while the interaction including $t_4$ was handled as a
perturbation.\par
%%%%%%%%%%%%%%%%%%%%%%%%%%%%%%%%%%%%%%%%%%%%%%%%%%%%%%%%%%%%%%%%%%%%%%%%%%%%%%%
The band structure of a phosphorene nanoribbon is shown in
Figure~\ref{Fig02}. A remarkable feature of a zPNR confined system
is the presence of the quasi-flat edge bands isolated from the bulk
modes. As is seen in~\ref{Fig02} these degenerate quasi-flat bands
in the middle of the conduction-valence gap entirely detached from
the bulk band. Adopting the above mentioned model, considering the
term including $t_4$ as a perturbation, the exact corresponding wave
function of such quasi-flat edge modes on an edge formed of atoms A,
can be written as~\cite{Ezawa2014}
%%%%%%%%%%%%%%%%%%%%%%%%%%%%%%%%%%%%%%%%%%%%%%%%%%%%%%%%%%%%%%%%%%%%%%%%%%%%%%%
\begin{equation}
|\Psi^A_k\rangle={1\over\sqrt{2\pi}}\sum_{m,n} \alpha^n(k)
\gamma(k)e^{ik(m+\delta_n)}|m,n\rangle. \label{EQ18}
\end{equation}
%%%%%%%%%%%%%%%%%%%%%%%%%%%%%%%%%%%%%%%%%%%%%%%%%%%%%%%%%%%%%%%%%%%%%%%%%%%%%%%
where $k$ is the wave-number. Here, without loss of generality, we
continue the discussion by considering an edge formed of $A$ atoms
and remark that the wave state on the sites of B atoms is zero.\par
%%%%%%%%%%%%%%%%%%%%%%%%%%%%%%%%%%%%%%%%%%%%%%%%%%%%%%%%%%%%%%%%%%%%%%%%%%%%%%%
In Eq.~(\ref{EQ18}), each lattice site is labeled by a pair of
integers $(m,n)$, where $m$ and $n$ are armchair and zigzag chain
numbers. It is assumed that for the considered edge $n=0$.\par
%%%%%%%%%%%%%%%%%%%%%%%%%%%%%%%%%%%%%%%%%%%%%%%%%%%%%%%%%%%%%%%%
The value of $\delta_n$ in Eq.~(\ref{EQ18}) is 0~(0.5) for
even~(odd) $n$, $\alpha(k)=-2\big(t_1/ t_2\big)\cos(k/2)$,  and
normalization factor $\gamma(k)$ satisfies the equation
$\gamma^2(k)=1-\alpha^2(k)$.\par
%%%%%%%%%%%%%%%%%%%%%%%%%%%%%%%%%%%%%%%%%%%%%%%%%%%%%%%%%%%%%%%%%%%%%%%%%%%%%%%%
It is obvious that the exact energy corresponding to the eigenstate
given in Eq.~(\ref{EQ18}) is zero. In a first-order approximation,
considering the perturbation interaction, it is straightforward to
show that eigenenergy corresponding to the above edge state changes
to~\cite{Ezawa2014}
%%%%%%%%%%%%%%%%%%%%%%%%%%%%%%%%%%%%%%%%%%%%%%%%%%%%%%%%%%%%%%%%%%%%%%%%%%%%%%%
\begin{equation}\begin{split}
E_k & = -4 \big(t_1t_4/ t_2\big)\big[1+\cos k\big]\\
    & = E_0-2t'\cos{k},
\end{split}
 \label{EQ19}
\end{equation}
%%%%%%%%%%%%%%%%%%%%%%%%%%%%%%%%%%%%%%%%%%%%%%%%%%%%%%%%%%%%%%%%%%%%%%%%%%%%%%%
where $E_0=-2 t' = -4 t_1 t_4/ t_2$ is an energy shift. As is seen,
energy shifts toward the negative values and the obtained dispersion
is very similar to what happens for a one-dimensional tight-binding
chain.
%%%%%%%%%%%%%%%%%%%%%%%%%%%%%%%%%%%%%%%%%%%%%%%%%%%%%%%%%%%%%%%%%%%%%%%%%%%%%%%
\begin{figure}[t]
\begin{center}
\includegraphics[scale=0.6]{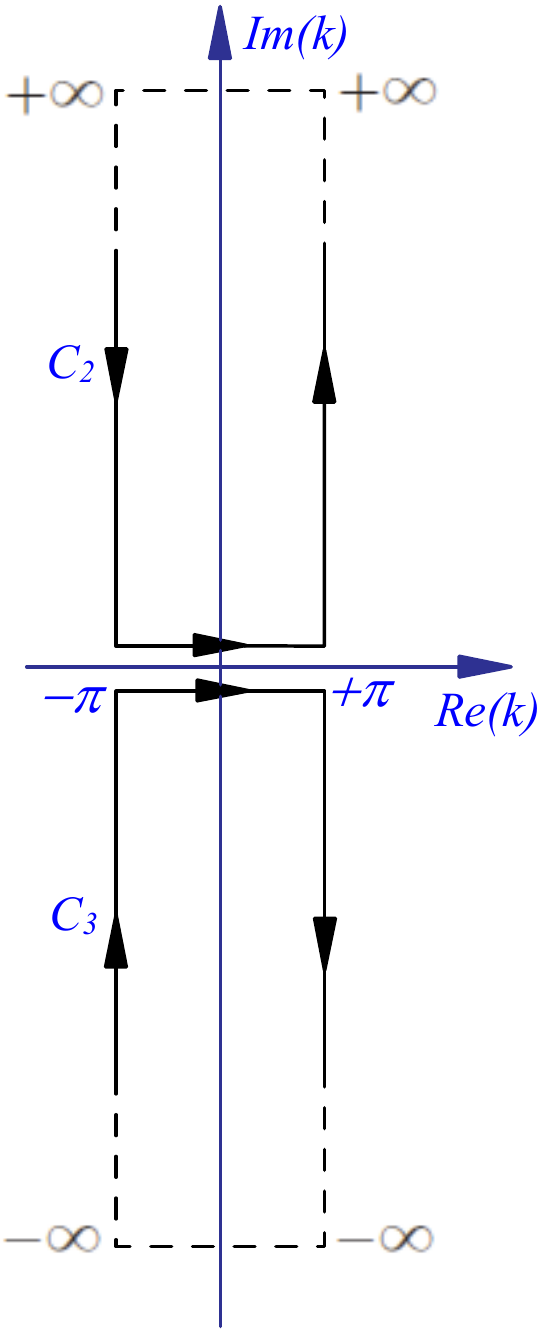}
\caption{(Color online) The integration contours in complex plane
used to derive the closed forms of the diagonal matrix elements of
Green operator ${\hat G}_E$, $C_2$ is used for $I_2$ and $C_3$ for
$I_3$.} \label{Fig03}
\end{center}
\end{figure}
%%%%%%%%%%%%%%%%%%%%%%%%%%%%%%%%%%%%%%%%%%%%%%%%%%%%%%%%%%%%%%%%%%%%%%%%%%%%%%%
\subsection{zPNR Green's function}
Scattering of the edge states by impurities is one of the main
issues to study in this section. To this end, since there exists a
considerable energy gap between the edge and bulk states, we need
only the Green function of the edge states in their energy domain.
The Green function corresponding to the edge states in zPNR can be
expressed in terms of the eigenstates $|\Psi_k^A\rangle$ and
eigenenergies $E_k$, respectively given in Eqs.~(\ref{EQ18})
and~(\ref{EQ19}), as
%%%%%%%%%%%%%%%%%%%%%%%%%%%%%%%%%%%%%%%%%%%%%%%%%%%%%%%%%%%%%%%%%%%%%%%%%%%%%%%
\begin{equation}
\label{EQ20} \hat{G}_E = \int_{-\pi}^\pi dk {|\Psi_k^A\rangle
\langle \Psi_k^A| \over E-E_k+i0^+}.
\end{equation}
%%%%%%%%%%%%%%%%%%%%%%%%%%%%%%%%%%%%%%%%%%%%%%%%%%%%%%%%%%%%%%%%%%%%%%%%%%%%%%%
The matrix elements of $\hat{G}_E$ can be evaluated as follows
%%%%%%%%%%%%%%%%%%%%%%%%%%%%%%%%%%%%%%%%%%%%%%%%%%%%%%%%%%%%%%%%%%%%%%%%%%%%%%%
\begin{equation}
\label{EQ21}
\begin{split}
G_E (m,n;m',n') & = \langle m,n|\hat{G}_E|m',n'\rangle\\
= {1\over 2\pi}\int_{-\pi}^\pi dk & {e^{-i(m- m'+\delta_n-\delta_n')
} \alpha^{n+n'}(k) \gamma^2(k) \over E-E_0+2 t' \cos k + i 0^+} .
\end{split}
\end{equation}
%%%%%%%%%%%%%%%%%%%%%%%%%%%%%%%%%%%%%%%%%%%%%%%%%%%%%%%%%%%%%%%%%%%%%%%%%%%%%%%
The typical integrals appearing in the above equation can be
evaluated by means of the residue theorem. For example, the diagonal
element of $m=n=m'=n'=0$ can be easily separated into two terms as
\begin{equation}\label{EQ22}
G_E(0,0;0,0) = [1-2(t_1/t_2)^2]I_1 - (t_1/t_2)^2(I_2+I_3),
\end{equation}
%%%%%%%%%%%%%%%%%%%%%%%%%%%%%%%%%%%%%%%%%%%%%%%%%%%%%%%%%%%%%%%%%%%%%%%%%%%%%%%
where
%%%%%%%%%%%%%%%%%%%%%%%%%%%%%%%%%%%%%%%%%%%%%%%%%%%%%%%%%%%%%%%%%%%%%%%%%%%%%%%
\begin{equation}
\label{EQ23}
\begin{split}
I_1 & = {1\over 2\pi} \int_{-\pi}^{\pi} dk {1\over E-E_0+2 t' \cos k +i0^+},\\
I_2 & ={1\over 2 \pi}  \int_{-\pi}^\pi dk {e^{ik} \over
E-E_0+2t'\cos k + i 0^+},\\
I_3 & ={1\over 2 \pi}  \int_{-\pi}^\pi dk {e^{-ik} \over
E-E_0+2t'\cos k + i 0^+}.
\end{split}
\end{equation}
%%%%%%%%%%%%%%%%%%%%%%%%%%%%%%%%%%%%%%%%%%%%%%%%%%%%%%%%%%%%%%%%%%%%%%%%%%%%%%%
For the first integral, $I_1$, we can use the variable change of
$z=e^{ik}$ to set $\cos k=(z+z^{-1})/2$ and  $dk=idz/z$. By doing
so, the denominator appears in the form of a second-order expression
in terms of $z$ with two solutions. These solutions are the simple
poles of the integrand. By closing the integration contour with a
unit radius circle around the origin of the complex plane, only one
of the poles occurs inside the contour. Employing the residue
theorem, the derivation results in
%%%%%%%%%%%%%%%%%%%%%%%%%%%%%%%%%%%%%%%%%%%%%%%%%%%%%%%%%%%%%%%%%%%%%%%%%%%%%%%
\begin{equation}
\label{EQ24} I_1 ={1/ 2 i t'  \sin k_0},
\end{equation}
%%%%%%%%%%%%%%%%%%%%%%%%%%%%%%%%%%%%%%%%%%%%%%%%%%%%%%%%%%%%%%%%%%%%%%%%%%%%%%%
where $\cos k_0=(E-E_0) / 2 t'$.\par
%%%%%%%%%%%%%%%%%%%%%%%%%%%%%%%%%%%%%%%%%%%%%%%%%%%%%%%%%%%%%%%%%%%%%%%%%%%%%%%
The integrand in $I_2$ has two simple poles at $k_1=k_0+i0^+$ and
$k_2=-k_0-i0^+$ where $k_0=\cos^{-1}[(E-E_0)/ 2t']$ with the same
residues of $e^{+ik_0}/ 2 t' \sin k_0$. In order to perform the
integral, we complete the integration contour by an infinite
rectangle in the upper half-plane as shown in Fig.~\ref{Fig03}. The
integrand over this contour vanishes as $Im(k)\to +\infty $. Also
the contribution of the vertical paths to the integral is zero for
the periodicity of the integrand. In this case, point $k_1$ is
within and $k_2$ is exterior to the contour. Consequently, the final
result for $I_2$ is
\begin{equation}
\label{EQ25} I_2 = e^{+ik_0}/ 2 i t' \sin k_0.
\end{equation}
Proceeding in a similar manner, we may also derive an analytical
form for $I_3$. In this case, since the exponential in the integrand
is negative, the contour should be completed by an infinite
rectangle in the lower half-plane as shown in Fig.~\ref{Fig03}. In
this case, the integral vanishes as $Im(k)\to -\infty $ and only
$k_2$ occurs inside the contour. Given these points, the result is
as expected equal to what we obtained for $I_2$.\par
%%%%%%%%%%%%%%%%%%%%%%%%%%%%%%%%%%%%%%%%%%%%%%%%%%%%%%%%%%%%%%%%%%%%%%%%%%%%%%%
The substitution of the derived closed-form expressions for $I_1$,
$I_2$ and $I_3$ into Eq.~(\ref{EQ22}) leads to
%%%%%%%%%%%%%%%%%%%%%%%%%%%%%%%%%%%%%%%%%%%%%%%%%%%%%%%%%%%%%%%%%%%%%%%%%%%%%%%
\begin{equation}\label{EQ26}
G_E(0,0;0,0) = {\gamma^2(k_0) \over 2 i t' \sin k_0}- {\big({t_1 /
t_2}\big)^2 \over 4 t'}.
\end{equation}
%%%%%%%%%%%%%%%%%%%%%%%%%%%%%%%%%%%%%%%%%%%%%%%%%%%%%%%%%%%%%%%%%%%%%%%%%%%%%%%
Following the same analysis, it is straightforward to obtain the
analytical form of the off-diagonal element $G_E(m,0;0,0)$ as
%%%%%%%%%%%%%%%%%%%%%%%%%%%%%%%%%%%%%%%%%%%%%%%%%%%%%%%%%%%%%%%%%%%%%%%%%%%%%%%
\begin{equation}\label{EQ27}
G_E(m,0;0,0) = {\gamma^2(k_0)e^{ik_0m} \over 2i t'\sin k_0}.
\end{equation}
The derived expressions~(\ref{EQ26}) and~(\ref{EQ27}) will be used
to study the scattering-induced entanglement in a zPNR.
%%%%%%%%%%%%%%%%%%%%%%%%%%%%%%%%%%%%%%%%%%%%%%%%%%%%%%%%%%%%%%%%%%%%%%%%%%%%%%%
%%%%%%%%%%%%%%%%%%%%%%%%%%%%%%%%%%%%%%%%%%%%%%%%%%%%%%%%%%%%%%%%%%%%%%%%%%%%%%%
%%%%%%%%%%%%%%%%%%%%%%%%%%%%%%%%%%%%%%%%%%%%%%%%%%%%%%%%%%%%%%%%%%%%%%%%%%%%%%%
\begin{figure*}[t]
\begin{center}
\includegraphics[scale=0.8]{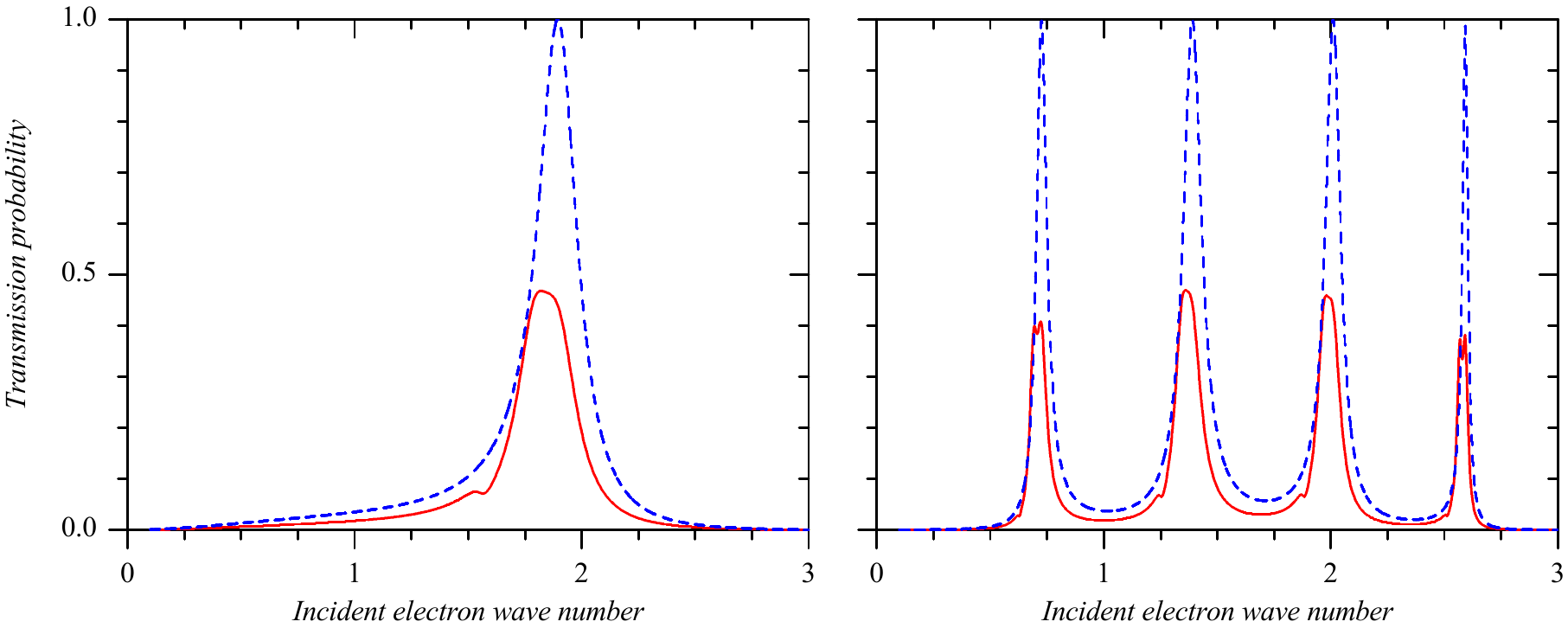}
\caption{(Color online) Transmission probability as a function of
the incident electron wave number for scattering from two spin
impurities doped into two sites of a one-dimensional tight-binding
chain. Solid~(red) and dashed~(blue) lines are respectively for the
initial states of
$|\hspace{-1mm}\uparrow\downarrow\downarrow\rangle$ and
$|\hspace{-1mm}\uparrow\uparrow\uparrow\rangle$. The left~(right)
panel is for $m=2~(m=5)$. For all the cases U'=10.} \label{Fig04}
\end{center}
\end{figure*}
%%%%%%%%%%%%%%%%%%%%%%%%%%%%%%%%%%%%%%%%%%%%%%%%%%%%%%%%%%%%%%%%%%%%%%%%%%%%%%%
%%%%%%%%%%%%%%%%%%%%%%%%%%%%%%%%%%%%%%%%%%%%%%%%%%%%%%%%%%%%%%%%%%%%%%%%%%%%%%%
%%%%%%%%%%%%%%%%%%%%%%%%%%%%%%%%%%%%%%%%%%%%%%%%%%%%%%%%%%%%%%%%%%%%%%%%%%%%%%%
%%%%%%%%%%%%%%%%%%%%%%%%%%%%%%%%%%%%%%%%%%%%%%%%%%%%%%%%%%%%%%%%%%%%%%%%%%%%%%%
%%%%%%%%%%%%%%%%%%%%%%%%%%%%%%%%%%%%%%%%%%%%%%%%%%%%%%%%%%%%%%%%%%%%%%%%%%%%%%%
%%%%%%%%%%%%%%%%%%%%%%%%%%%%%%%%%%%%%%%%%%%%%%%%%%%%%%%%%%%%%%%%%%%%%%%%%%%%%%%
%%%%%%%%%%%%%%%%%%%%%%%%%%%%%%%%%%%%%%%%%%%%%%%%%%%%%%%%%%%%%%%%%%%%%%%%%%%%%%%
%%%%%%%%%%%%%%%%%%%%%%%%%%%%%%%%%%%%%%%%%%%%%%%%%%%%%%%%%%%%%%%%%%%%%%%%%%%%%%%
\subsection{Impurity entanglement through electron scattering}
In this subsection, we assume that two spin impurities are doped
into two edge sites of an  $A$-type zigzag chain of a phosphorene
nanoribbon. The edge-sate electrons traveling along the zigzag chain
scatter off the impurities. We will show that the spin interaction
between the incident electrons and the impurities during the
scattering process causes the entanglement of the final spin state
of the specified system. As is understood from the above subsections
and also is seen in figure~\ref{Fig04}, the situation is very
similar to what we discussed for a one-dimensional tight-binding
chain in section~\ref{Sec02}. In this case, by labeling the impurity
sites by $(0,0)$ and $(m,0)$, the explicit form of the interaction
due to their presence is
%%%%%%%%%%%%%%%%%%%%%%%%%%%%%%%%%%%%%%%%%%%%%%%%%%%%%%%%%%%%%%%%%%%%%%%%%%%%%%%
\begin{equation}
\label{EQ28} \hat{V} = U \big[ ({\bf S}_1 \cdot {\bf S}_2)
c_{0,0}^\dagger c_{0,0} + ({\bf S}_1 \cdot {\bf S}_3)
c^\dagger_{m,0} c_{m,0} \big],
\end{equation}
%%%%%%%%%%%%%%%%%%%%%%%%%%%%%%%%%%%%%%%%%%%%%%%%%%%%%%%%%%%%%%%%%%%%%%%%%%%%%%%
where $c_{0,0}^\dagger$ and $c^\dagger_{m,0}$ ($c_{0,0}$ and
$c_{m,0}$) are the creation~(annihilation) operators of electrons in
sites $(0,0)$ and $(m,0)$, respectively, and the other quantities
are the same as introduced previously. The incoming state is assumed
for example as
\begin{equation}
\label{EQ29} | \Psi_{in} \rangle =  |\Psi_{k}^{A} \rangle
|\hspace{-1mm}\uparrow\downarrow\downarrow\rangle.
\end{equation}
Using the Lippmann–Schwinger equation and following the approach
explained in section~\ref{Sec02}, the reflected and transmitted wave
states are respectively given by
%%%%%%%%%%%%%%%%%%%%%%%%%%%%%%%%%%%%%%%%%%%%%%%%%%%%%%%%%%%%%%%%%%%%%%%%%%%%%%%
\begin{equation}
\label{EQ30} |\Psi_R \rangle  = |\Psi_{-k}^{A}
\rangle|S_R\rangle;\qquad {\rm~for} \ m'<0,
\end{equation}
and
\begin{equation}
\label{EQ31} |\Psi_T \rangle  = |\Psi_{k}^{A} \rangle
|S_T\rangle;\qquad {\rm~for} \ m'>m,
\end{equation}
%%%%%%%%%%%%%%%%%%%%%%%%%%%%%%%%%%%%%%%%%%%%%%%%%%%%%%%%%%%%%%%%%%%%%%%%%%%%%%%
with the reflected and transmitted spin states of
\begin{equation}
\label{EQ32}
\begin{split}
|S_R\rangle & = [G_E(m',0;0,0) | s_0 \rangle\\ &\hspace{12mm} + G_E(m',0;m,0) |s_m \rangle]e^{+ik_0m'},\\
|S_T\rangle & =|\hspace{-1mm}\uparrow\downarrow\downarrow\rangle +
[G_E(m',0;0,0) |s_0\rangle\\ &\hspace{12mm}  + G_E(m',0;m,0) |s_m
\rangle] e^{-ik_0m'}.
\end{split}
\end{equation}
Here, also similar to the one-dimensional case discussed in
section~\ref{Sec02}, the total spin states of $|s_0\rangle$ and
$|s_m\rangle$ are exactly known and their explicit forms can be
derived analytically.\par
%%%%%%%%%%%%%%%%%%%%%%%%%%%%%%%%%%%%%%%%%%%%%%%%%%%%%%%%%%%%%%%%%%%%%%%%%%%%%%%
The wave states given in equations~(\ref{EQ29}) and~(\ref{EQ30}) can
be used to the obtain the relevant partial density matrices and
those in turn may be used to calculate the negativity as a measure
of the produced entanglement in the final spin state of the system.
Also, the reflection and transmission probabilities are computable
using the above quantum states.\par
%%%%%%%%%%%%%%%%%%%%%%%%%%%%%%%%%%%%%%%%%%%%%%%%%%%%%%%%%%%%%%%%%%%%%%%%%%%%%%%
The calculations in this and previous sections can be repeated for
any given initial spin state.
%%%%%%%%%%%%%%%%%%%%%%%%%%%%%%%%%%%%%%%%%%%%%%%%%%%%%%%%%%%%%%%%%%%%%%%%%%%%%%%
%%%%%%%%%%%%%%%%%%%%%%%%%%%%%%%%%%%%%%%%%%%%%%%%%%%%%%%%%%%%%%%%%%%%%%%%%%%%%%%
\begin{figure*}
\centering
\includegraphics[scale=0.8]{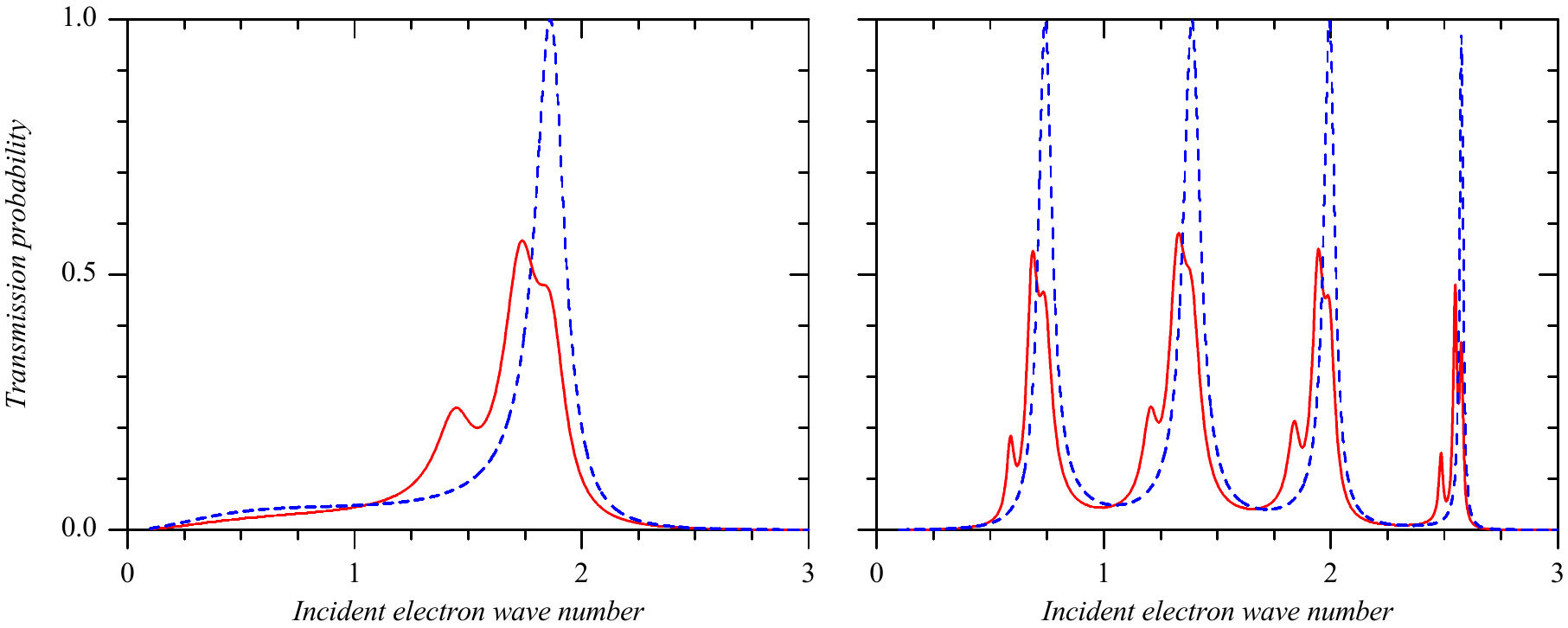}
\caption{Same as figure~\ref{Fig04}, but for scattering of the
edge-state electrons from two impurities doped into two sites of an
A-type zPNR.} \label{Fig05}
\end{figure*}
%%%%%%%%%%%%%%%%%%%%%%%%%%%%%%%%%%%%%%%%%%%%%%%%%%%%%%%%%%%%%%%%%%%%%%%%%%%%%%%
%%%%%%%%%%%%%%%%%%%%%%%%%%%%%%%%%%%%%%%%%%%%%%%%%%%%%%%%%%%%%%%%%%%%%%%%%%%%%%%
\begin{figure}
\centering
\includegraphics[scale=0.8]{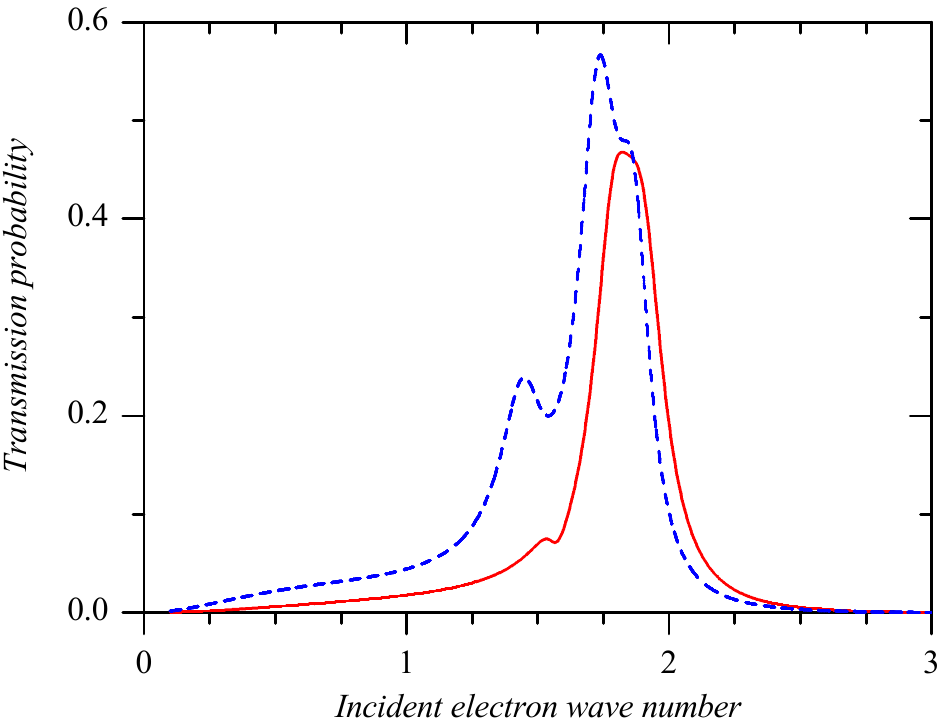}
\caption{Comparison of the transmission probabilities for scattering
of free electrons from impurities doped into two sites of a one
dimensional chain~(solid line) and scattering of the edge-state
electrons from a zigzag chain in a phosphorene nanoribbon~(dashed
line). For both cases the initial spin state is
$|\hspace{-1mm}\uparrow\downarrow\downarrow\rangle$, $m=0$ and
$U'=10$.}\label{Fig06}
\end{figure}
%%%%%%%%%%%%%%%%%%%%%%%%%%%%%%%%%%%%%%%%%%%%%%%%%%%%%%%%%%%%%%%%%%%%%%%%%%%%%%%
%%%%%%%%%%%%%%%%%%%%%%%%%%%%%%%%%%%%%%%%%%%%%%%%%%%%%%%%%%%%%%%%%%%%%%%%%%%%%%%
\begin{figure*}
\centering
\includegraphics[scale=0.8]{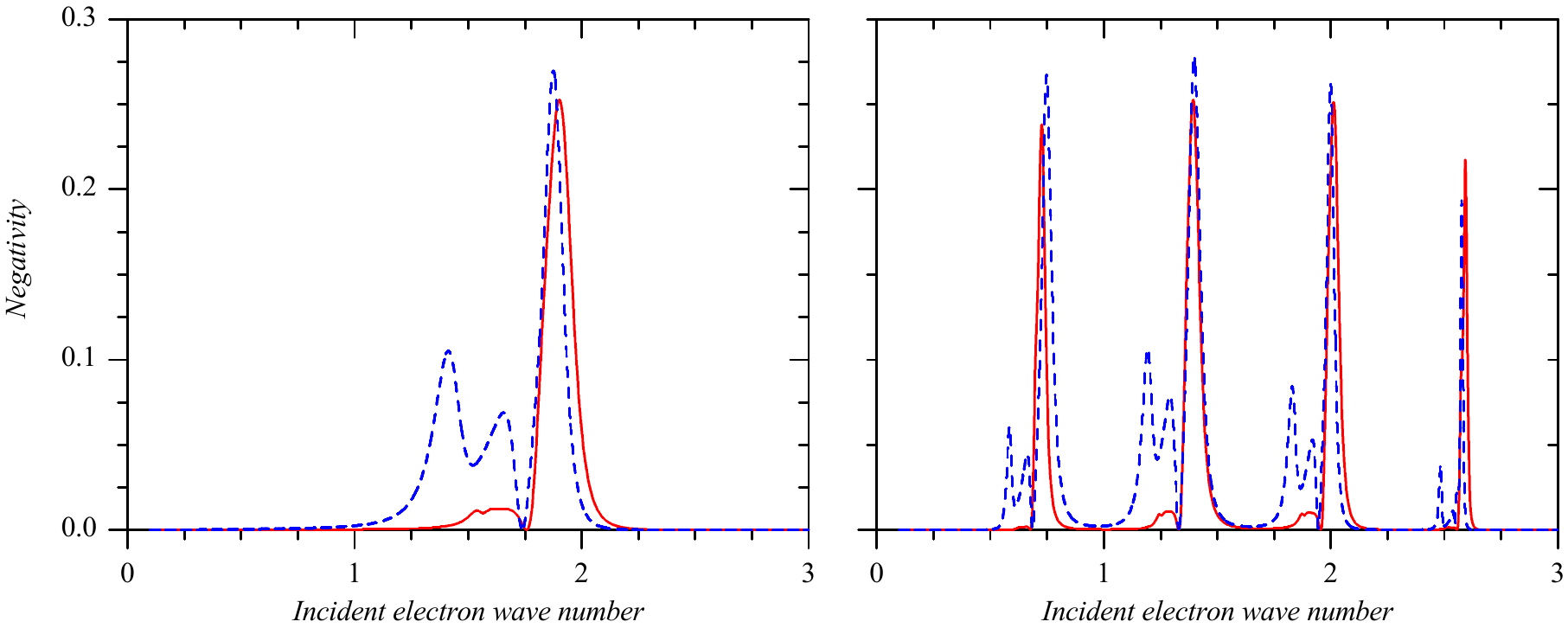}
\caption{Negativity vs the incident electron wave number as a
measure of the created entanglement due to the scattering of the
free~(edge-state) electrons from two spin impurities doped into two
sites of a one-dimensional tight-binding chain (an edge zigzag chain
of phospherene). Solid~(red) and dashed~(blue) lines are
respectively for chain and phosphorene. The initial spin state in
all the cases is
$|\hspace{-1mm}\uparrow\downarrow\downarrow\rangle$. The
left~(right) panel is for $m=2~(m=5)$. For all the cases $U'=10$.}
\label{Fig07}
\end{figure*}
%%%%%%%%%%%%%%%%%%%%%%%%%%%%%%%%%%%%%%%%%%%%%%%%%%%%%%%%%%%%%%%%%%%%%%%%%%%%%%%
%%%%%%%%%%%%%%%%%%%%%%%%%%%%%%%%%%%%%%%%%%%%%%%%%%%%%%%%%%%%%%%%%%%%%%%%%%%%%%%
\begin{figure}
\centering
\includegraphics[scale=0.8]{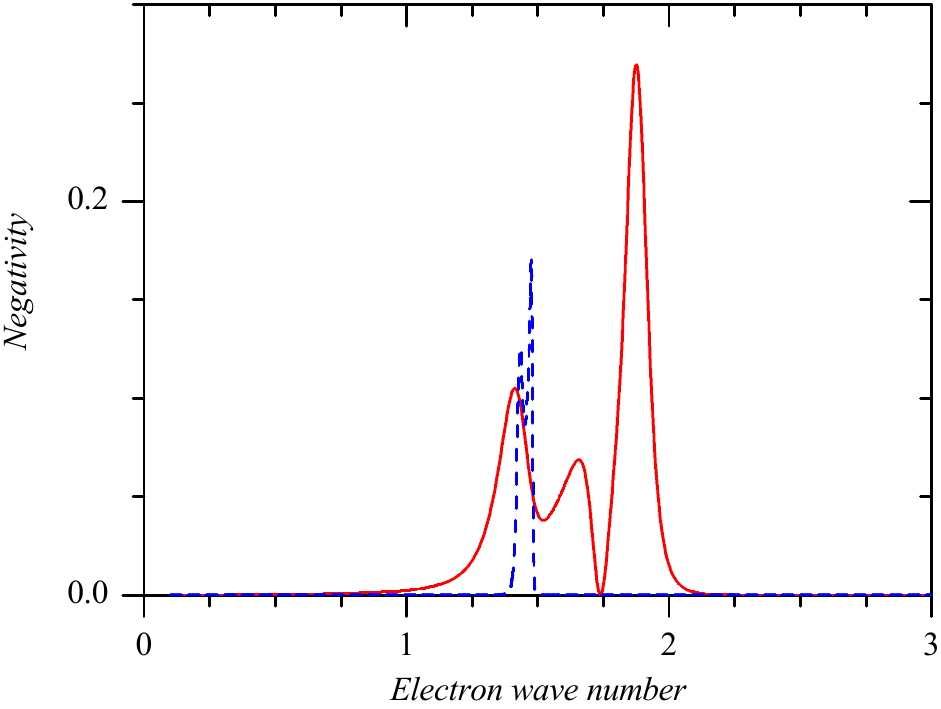}
\caption{The created entanglement between two on-side magnetic
impurities in an edge zigzag chain of phosphorene nanoribbon as a
function of the incident electrons wave number. The initial state is
$|\hspace{-1mm}\uparrow\downarrow\downarrow\rangle$, $m=2$, but
$U'=10$~(solid line) and $U'=100$~(dashed line).} \label{Fig08}
\end{figure}
%%%%%%%%%%%%%%%%%%%%%%%%%%%%%%%%%%%%%%%%%%%%%%%%%%%%%%%%%%%%%%%%%%%%%%%%%%%%%%%
%%%%%%%%%%%%%%%%%%%%%%%%%%%%%%%%%%%%%%%%%%%%%%%%%%%%%%%%%%%%%%%%%%%%%%%%%%%%%%%
%%%%%%%%%%%%%%%%%%%%%%%%%%%%%%%%%%%%%%%%%%%%%%%%%%%%%%%%%%%%%%%%%%%%%%%%%%%%%%%
%%%%%%%%%%%%%%%%%%%%%%%%%%%%%%%%%%%%%%%%%%%%%%%%%%%%%%%%%%%%%%%%%%%%%%%%%%%%%%%
\section{Results\label{Sec04}}
In this section, in order to demonstrate the performance of the
models discussed in the previous sections, several examples of the
electron scattering-induced entanglement between the doped magnetic
impurities are presented and discussed. In following discussions, it
is assumed that the impurities are localized at sites labeled by $0$
and $m$ in a one-dimensional tightly bonded chain, and at sites
$(0,0)$ and $(m,0)$ in the edge zigzag chain of a phosphorene
nanoribbon. So, with $m$, the second impurity location is completely
known in the both cases. Also, the strength of the scattering
potential, $U$, is normalized to $U'$, where $U'=U/t$ for chain and
$U'=U/t'$ for phosphorene with $t'=2t_1t_4/t_2$.\par\par
%%%%%%%%%%%%%%%%%%%%%%%%%%%%%%%%%%%%%%%%%%%%%%%%%%%%%%%%%%%%%%%%%%%%%%%%%%%%%%%
Figure~\ref{Fig04} presents the electron transmission probability
for scattering of free traveling electrons along a one-dimensional
tight-binding chain. The incident electrons scatter off two
localized impurities at sites $0$ and $m$. The results are shown for
$m=2$ and $m=5$ with initial spin states of
$|\hspace{-1mm}\uparrow\downarrow\downarrow\rangle$ and
$|\hspace{-1mm}\uparrow\uparrow\uparrow\rangle$. Strength of the
normalized scattering potential, $U'$, is set equal to 10.\par
%%%%%%%%%%%%%%%%%%%%%%%%%%%%%%%%%%%%%%%%%%%%%%%%%%%%%%%%%%%%%%%%%%%%%%%%%%%%%%%
As is seen from the figure, for the initial spin state in which all
the three spins are in same direction, the behavior of the
transmission probability is similar to what happens for scattering
due to the on-site spinless interactions. In  this case, for several
values of the incident electron wave numbers, resonance occurs. The
number of the resonance peaks are $m-1$ and their hight are equal to
unit. With changing $U'$ the number of the resonance peaks and their
hight remain unchanged.\par
%%%%%%%%%%%%%%%%%%%%%%%%%%%%%%%%%%%%%%%%%%%%%%%%%%%%%%%%%%%%%%%%%%%%%%%%%%%%%%%
Also, for the initial spin state of
$|\hspace{-1mm}\uparrow\downarrow\downarrow\rangle$, in which one of
the spins is in opposite direction of the others, several resonance
peaks are observed in the transition probability, but the hight of
the peaks are smaller than that of the corresponding peaks for
$|\hspace{-1mm}\uparrow\uparrow\uparrow\rangle$. For the initial
spin state of $|\hspace{-1mm}\uparrow\downarrow\downarrow\rangle$,
the electron transmission probability is a bit dependent on $U'$
value.\par
%%%%%%%%%%%%%%%%%%%%%%%%%%%%%%%%%%%%%%%%%%%%%%%%%%%%%%%%%%%%%%%%%%%%%%%%%%%%%%%
%%%%%%%%%%%%%%%%%%%%%%%%%%%%%%%%%%%%%%%%%%%%%%%%%%%%%%%%%%%%%%%%%%%%%%%%%%%%%%%
A similar situation is investigated for scattering of the edge-state
electrons from two on-site doped impurities into a  A-type edge
zigzag chain in a phosphorene nanoribbon. For this case, the changes
of the electron transmission probability in terms of the incident
electron wave number are presented in Fig.~\ref{Fig05}. As is seen
from the figure, the behavior exhibited in this case is similar to
what was observed for scattering from a one-dimensional chain. For
the initial spin state with three spins in the same direction, the
hight of the resonance peaks is unit but for other initial spin
states this hight reduces considerably. The number of the resonance
peaks in this case is also $m-1$. These facts confirm our assertion
that the zigzag edge chains in phosphorene nanoribbons  behaves like
a one-dimensional tight-binding chain.\par
%%%%%%%%%%%%%%%%%%%%%%%%%%%%%%%%%%%%%%%%%%%%%%%%%%%%%%%%%%%%%%%%%%%%%%%%%%%%%%%
%%%%%%%%%%%%%%%%%%%%%%%%%%%%%%%%%%%%%%%%%%%%%%%%%%%%%%%%%%%%%%%%%%%%%%%%%%%%%%%
In figure~\ref{Fig06}, the changes of the electron transmission
probabilities in terms of the the incident electron wave number for
scattering of the free traveling electrons from the impurities in a
one-dimensional tight-binding chain and for scattering of edge-state
traveling electrons from the impurities in a phosphorene zigzag
chain are compared. For both considered cases, $m=2$, $U'=10$ and
the initial spin state is assumed as
$|\hspace{-1mm}\uparrow\downarrow\downarrow\rangle$. As is seen the
graphs are very similar in overall futures, but the resonance peak
for phosphorene is a little higher than that for chain. This is for
the fact that the impurities in phosphorene are fixed on the edge
sites, while the edge-state wave function  slightly penetrates into
the bulk.\par
%%%%%%%%%%%%%%%%%%%%%%%%%%%%%%%%%%%%%%%%%%%%%%%%%%%%%%%%%%%%%%%%%%%%%%%%%%%%%%%
%%%%%%%%%%%%%%%%%%%%%%%%%%%%%%%%%%%%%%%%%%%%%%%%%%%%%%%%%%%%%%%%%%%%%%%%%%%%%%%
The electron scattering leads to entanglement production between the
spin impurities in the both specified cases. Negativity obtained
using the partial density matrix $\rho_{23}$ is a measure of the
produced entanglement. This measure is displayed as a function of
the electron wave number in Fig.~\ref{Fig07}, for scattering of
electrons in a chain and in a phosphorene nanoribbon. For all the
cases, the initial state  is
$|\hspace{-1mm}\uparrow\downarrow\downarrow\rangle$ and $U'=10$. The
results are shown for $m=2$ and $m=5$.\par
%%%%%%%%%%%%%%%%%%%%%%%%%%%%%%%%%%%%%%%%%%%%%%%%%%%%%%%%%%%%%%%%%%%%%%%%%%%%%%%
Several points are remarkable from this figure. The resonance peaks
are observable for the produced entanglement. In addition to the
main resonance peaks, a number of the local small peaks are also
seen in the resonance spectrum of the created correlation. Structure
of these local peaks for phosphorene is more complex than that for
chain, also these peaks for phosphorene are considerably higher than
those of chain. This is for the fact that the edge-state wave
function for phosphorene is more complicated than the considered
wave function for chain. As previous, the number of the main
resonance peaks is $m-1$ for both chain and nanoribbon. Overall
aspects of the main peaks in phosphorene are similar to those in
chain, however the phosphorene main peaks are usually a little
higher than those of chain. In the resonance case the negativity
tends to the considerable value of 0.3. This shows that the present
model is very efficient in producing a considerable amount of
entanglement.\par
%%%%%%%%%%%%%%%%%%%%%%%%%%%%%%%%%%%%%%%%%%%%%%%%%%%%%%%%%%%%%%%%%%%%%%%%%%%%%%%
In figure~\ref{Fig08}, we investigated the dependence of the created
entanglement between the on-side impurities in the phosphorene
nanoribbon on the normalized strength of the interaction potential
$U'$. To this end, the negativity is plotted as a function of the
incident electron wave number for two values of $U'$; $U'=10$ and
$U'=100$. As is seen, with increasing $U'$, the local peaks
disappear and the main peaks becomes more sharper but their hight
decreases.  A similar behavior is also observable for the electron
transmission probability. In fact, the increase in $U'$ causes that
the electron transmission probability becomes considerable only at
certain specific resonance energies. As a result the produced
entanglement between the impurities is significant only at theses
resonance situations.\par
%%%%%%%%%%%%%%%%%%%%%%%%%%%%%%%%%%%%%%%%%%%%%%%%%%%%%%%%%%%%%%%%%%%%%%%%%%%%%%%
%%%%%%%%%%%%%%%%%%%%%%%%%%%%%%%%%%%%%%%%%%%%%%%%%%%%%%%%%%%%%%%%%%%%%%%%%%%%%%%
%%%%%%%%%%%%%%%%%%%%%%%%%%%%%%%%%%%%%%%%%%%%%%%%%%%%%%%%%%%%%%%%%%%%%%%%%%%%%%%
%%%%%%%%%%%%%%%%%%%%%%%%%%%%%%%%%%%%%%%%%%%%%%%%%%%%%%%%%%%%%%%%%%%%%%%%%%%%%%%
%%%%%%%%%%%%%%%%%%%%%%%%%%%%%%%%%%%%%%%%%%%%%%%%%%%%%%%%%%%%%%%%%%%%%%%%%%%%%%%
\section{Summary and Conclusions \label{Sec05}}
%%%%%%%%%%%%%%%%%%%%%%%%%%%%%%%%%%%%%%%%%%%%%%%%%%%%%%%%%%%%%%%%%%%%%%%%%%%%%%%
We studied the scattering of the edge-state electrons, traveling
along an edge zigzag chain of a phosphorene nanoribbon, from two
magnetic impurities localized at two sites of this chain. It was
shown that the situation is very similar to the scattering of free
traveling electrons from a one-dimensional tight-binding chain with
two on-side spin impurities. With a given initial wave state, the
Lippmann-Schwinger equation, the tight-binding model  and the
Green's function approach were employed to calculate the outgoing
wave state, analytically. Using the provided model, the reflected
and transmitted parts of the final wave state and their relevant
partial density matrices were derived. Consequently, the reflection,
and transmission probabilities and the negativity as a measure of
the created entanglement between the impurities were calculated.
Several examples were presented and discussed to show the
performance of the suggested model. It was shown that, for certain
resonance energies, both the electron transmission probability and
the generated correlation between the magnetic impurities are
considerable. The importance of the performed research is that it
proposed a method for creating the entanglement between two magnetic
impurities, which possibly can be realized using the experimental
methods.\par
%%%%%%%%%%%%%%%%%%%%%%%%%%%%%%%%%%%%%%%%%%%%%%%%%%%%%%%%%%%%%%%%%%%%%%%%
\begin{acknowledgments}
The forth author would like to acknowledge the office of graduate
studies at the University of Isfahan for their support and research
facilities
\end{acknowledgments}
%%%%%%%%%%%%%%%%%%%%%%%%%%%%%%%%%%%%%%%%%%%%%%%%%%%%%%%%%%%%%%%%%%%%%%%%
%%%%%%%%%%%%%%%%%%%%%%%%%%%%%%%%%%%%%%%%%%%%%%%%%%%%%%%%%%%%%%%%%%%%%%%%%%%%%%%

%%%%%%%%%%%%%%%%%%%%%%%%%%%%%%%%%%%%%%%%%%%%%%%%%%%%%%%%%%%%%%%%%%%%%%%%%%%%%%%

\end{document}